\documentclass[pre,citeautoscript,superscriptaddress,twocolumn]{revtex4}

\usepackage{amsfonts,amsmath,amssymb}
\usepackage[dvips]{graphicx,color}

\begin{document}

\title{Stabilization of Membrane Pores by Packing}

\author{D. J. Bicout}
\affiliation{Institut Laue-Langevin, 6 rue Jules Horowitz,
B.P. 156, 38042 Grenoble, France}
\affiliation{Biomathematics and Epidemiology, ENVL - TIMC, B.P. 83,
69280 Marcy l'Etoile, France}

\author{F. Schmid}
\affiliation{Institut Laue-Langevin, 6 rue Jules Horowitz,
B.P. 156, 38042 Grenoble, France}
\affiliation{Fakult\"at f\"ur Physik, Universit\"at Bielefeld,
D-33615 Bielefeld, Germany}

\author{E. Kats}
\affiliation{Institut Laue-Langevin, 6 rue Jules Horowitz,
B.P. 156, 38042 Grenoble, France}
\affiliation{L. D. Landau Institute for Theoretical Physics,
RAS, 117940 GSP-1, Moscow, Russia}


\begin{abstract}
We present a model for pore stabilization in membranes without surface tension.
Whereas an isolated pore is always unstable (since it either shrinks tending 
to re-seal or grows without bound til to membrane disintegration), it is shown 
that excluded volume interactions in a system of many pores can stabilize 
individual pores of a given size in a certain range of model parameters. For 
such a multipore membrane system, the distribution of pore size and associated 
pore lifetime are calculated within the mean field approximation. We predict 
that, above the temperature ${\rm T_m}$ when the effective line tension 
becomes negative, the membrane exhibits a dynamic sieve-like porous structure.
\end{abstract}

\maketitle

\begin{figure}[tb]
\includegraphics[width=0.4\textwidth,angle=0]{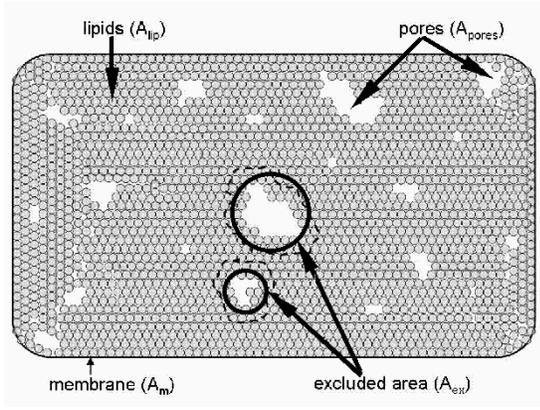} 
\caption{Schematic depiction of a porous membrane, illustrating the 
different areas introduced in the text: The total surface $A_{\rm lip}$ 
covered by all lipids (gray filled circles), the total area of
pores $A_{\rm pores}$, and the total area 
$A_{\rm m}=A_{\rm lip}+A_{\rm pores}$.
The circle around the pore represents the area that is effectively
excluded by the pore, and the total excluded area for all pores 
is $A_{\rm ex}$. Due to the fractal shape of pores, the excluded
area of pores is larger than their actual area.}
 \label{fig1}
\end{figure}

Pores can form and grow in membranes in response to thermal
fluctuations and external influences. Pore growth enhances the
transport of biomolecules across the membranes and its biological
relevance can bring new prospective biotechnological applications
(see, e.g., Ref.~\cite{AA79,SK01}). Holes appear in the membrane
via a thermally activated poration process, and their subsequent
growth is controlled by the effective line tension (assuming a
negligible small surface tension). A schematic depiction of a 
membrane with holes is displayed in Fig.\ref{fig1}. Preliminary
computer simulations of coarse-grained models for lipid membranes
have indicated that close to their disintegration point, membranes
sometimes exhibit a perforated state where they are peppered
of many pores~\cite{cooke_poster}. In the present paper, we 
investigate a mechanism that can stabilize such a state.

We first consider a single pore formation in a membrane with
zero surface tension. For simplicity, the pore edge is regarded as
a closed self-avoiding path of $n$ steps of constant length $l$.
In the simplest approximation, the free energy of the system
(tensionless membrane with a pore of size $n$) at nonzero temperature 
can be written as a summation of two contributions: a purely energetic
part as suggested by Litster \cite{LI75}, plus an entropic part as
modeled by Shillcock and Boal \cite{SB96},
$f(n)=f_0+\lambda_0\,l\,n-{\rm k_BT}\ln[\omega(n)]$, 
where $f_0$ is a $n$-independent energy, $\lambda_0$ the bare line 
tension of the pore edge, ${\rm k_BT}=1/\beta$ the thermal energy, and
$\omega(n)$ the number of possible conformations of the pore
contour of size $n$. For self-avoiding walks, $\omega(n)$ has a
general form \cite{MK76,GL88} given by
$\omega(n)=\omega_0\,z^n\,n^{\alpha-2}$, where $\omega_0$ is a
constant, $z$ the connectivity constant of the medium, and the 
exponent $\alpha=1/2$ for self-avoiding random walks 
in two dimensions and $\alpha < 2$ for any kind of pores. Let
$n_0$ be the minimal size of the pore, the above free energy 
can be rewritten as $f(n)=0$ for $0\leq n\leq n_0$, and for $n>n_0$, 
\begin{eqnarray}\label{b1} 
f(n)&=&
F_0+\overbrace{\lambda_0\,\left(1-\frac{\rm T}{\rm T_m}\right)}^{\lambda_1}
\,l\,\left[n-n_0\right]\nonumber 
\\
&&\quad +\: (2-\alpha)\,{\rm k_BT}\ln\left(\frac{n}{n_0}\right)\:,
\end{eqnarray}
where we have defined $F_0$ as the free energy required to create
or initiate a minimal-sized pore, $\lambda_1$ the entropically modified 
line tension of the pore edge, and ${\rm T_m}=\lambda_0\,l/
{\rm k_B}\ln(z)$ the disintegration temperature. In what follows, 
we will neglect the logarithmic term, $(2-\alpha)\ln(n/n_0)$, 
that only slightly renormalizes the results. Simple inspection 
of Eq.(\ref{b1}) indicates that the free energy of the system monotonically
increases as the pore grows larger at low ${\rm T}<{\rm T_m}$ when 
$\lambda_1>0$, keeping the membrane stable with 
an unstable pore that reseals. In this case, a pore of any size has 
a finite lifetime and will ultimately shrinks to disappearance. In 
contrast, when the effective line tension is negative, $\lambda_1<0$, 
at high ${\rm T}>{\rm T_m}$, all newly initiated pores 
grow without bound (i.e., with a diverging lifetime) leading hence 
to destabilization and disintegration of the membrane. There are
numbers of works that have investigated mechanisms for stabilizing
membranes with a single pore. These include, for instance,
membrane bending fluctuations, renormalization of linear and
surface tension coefficients \cite{FS05}, area exchange in tense
membranes \cite{SSa98}, osmotic stress \cite{IL04}, hydrodynamics
\cite{MN97,ZN93}, orientational ordering \cite{GP96}, and others
(see Ref.~\cite{LS95} for more details and references). In this paper 
we focus on a new stabilizing effect, which is created by the 
presence of multiple pores in the membrane.

Indeed, as the probability of initiating several pores on a membrane 
increases as the temperature gets higher, we are now confronted to the 
situation of a membrane with an ensemble of pores (as in Ref.~\cite{LM04}). 
If the bare line tension is negative, the system tries to create as
much pore rim as possible~\cite{fn_membrane}. In that case, a membrane 
state with many small pores is more favorable than one with only one very 
large pore. In some sense, such a state is similar to a droplet 
microemulsion structure in amphiphilic systems, where the fluid is 
macroscopically homogeneous, but filled with internal interfaces on 
the microscopic scale~\cite{Gompper}. 
To quantify this expectation, we proceed as follows: We first consider 
the free energy $F\{N(n)\}$ of a membrane where the number $N(n)$ of 
pores with contour length $n$ is fixed (constrained). In the second step, 
we will relieve the constraint and minimize $F\{N(n)\}$ with respect 
to $N(n)$. The free energy $F\{N(n)\}$ has energetic line tension
contributions and entropic shape contributions as in Eq.~(\ref{b1}). 
Moreover, the pores have the translational entropy of a two dimensional
gas. However, they may not overlap, since a configuration with
two ``overlapping'' pores would have to be replaced by a new
configuration with just one, larger pore, and a different $N(n)$.
For fixed $N(n)$, this restricts the translational degree of freedom 
of pores, as if they had excluded volume interactions. 
Note that these ``interactions'' are purely entropic. One could also 
introduce real repulsive interactions between pores, originating, e.g., 
from direct electrostatic or van der Waals forces, entropic or Helfrich 
interactions arising from pore shape fluctuations, or from membrane 
undulations \cite{GB93,MM02}. However, this is not necessary for
our argument.

To proceed we adopt a simple van der Waals approach and approximate
the free energy $F\{N(n)\}$ by
$F\{N(n)\}=\sum_{n}N(n)f(n)-{\rm k_BT}\,\ln\left[\Omega\right]$, 
where $f(n)$ is the free energy of a single pore as given in 
Eq.(\ref{b1}), and the total translational entropy $\Omega$ of the pores 
is given by
\begin{equation}\label{om}
\begin{array}{rcl}
\Omega&=&\displaystyle{\prod_{n}\frac{1}{[N(n)]!}\,
\left[\frac{A_{\rm m}-A_{\rm ex}}{a_0}\right]^{N(n)}}
\\
&\approx &
\displaystyle{\prod_{n}\left[\frac{(A_{\rm lip}-A_0)\,e}{a_0\,N(n)}\right]^{N(n)}}.
\end{array}
\end{equation}
Here $a_0$ is an area constant defined below, $A_{\rm m}$ is the total membrane area, 
$A_{\rm m}=A_{\rm lip}+A_{\rm pores}$, with the area of lipids $A_{\rm lip}$
and the pore area $A_{\rm pores}$, and $A_{\rm ex}$ is the area that is 
effectively inaccessible to a test pore due to the presence of the other
pores (the excluded area). Due to the fractal nature of the pores, $A_{\rm ex}$ 
is larger than $A_{\rm pores}$ as indicated in Fig.~\ref{fig1}. 
Notice that $A_{\rm lip}$ is constant since it is proportional to the number of 
lipids. In writing the rightmost expression in Eq.(\ref{om}), we have used the 
approximation, $N!\approx N^N\,{\rm e}^{-N}$, and rewritten the accessible area 
for pores as $A_{\rm m}-A_{\rm ex}=A_{\rm lip}-A_0$, such that 
$A_0=A_{\rm ex}-A_{\rm pores}=\displaystyle\sum_{n}N(n)a(n)$, 
where $a(n)$ is the difference between the excluded and
the actual areas of a pore of size $n$. Since the contours of the pores
have self-avoiding walk statistics, $a(n)$ scales like $a(n) = a_0 (n/n_0)^{2 \nu}$
with the Flory exponent $\nu \approx 3/4$. This defines $a_0$. Now, inserting
Eq.(\ref{om}) into the free energy expression, and minimizing with respect 
to $N(n)$, yields the normalized equilibrium distribution $P_{\rm eq}(n)$ 
of pore sizes,
\begin{equation}
P_{\rm eq}(n)=N(n)\left/\sum_{n=1}^{\infty} N(n)\right.
=\frac{\exp\left\{-\beta G(n)\right\}}{Q}\:.
\end{equation}
Here we have defined the effective free energy,
$G(n)=f(n) + {\rm k_BT} Q (n/n_0)^{2 \nu}$. The partition
function $Q$ can be regarded as a packing density of pores, and is given 
self-consistently as,
\begin{eqnarray}\label{part}
Q&=&\frac{a_0}{(A_{\rm lip}-A_0)}\,\sum_{n}N(n)
=\int_{1}^{\infty}\!{\rm e}^{-\beta G(x)}\,dx\:,
\end{eqnarray}
Defining $x=n/n_0$ and using $\nu = 3/4$, the effective free energy 
$G(x)$ of a pore in a membrane containing an ensemble of pores 
(in gas phase) reads as $G(x)=0$ for $0\leq x<1$, and, 
\begin{eqnarray}\label{fener}
G(x)=F_0+Q{\rm k_BT}+n_0\,l\int_{1}^{x}\!\lambda_{\rm eff}(z)\,dz\:\:;\:\:x\geq 1\:,
\end{eqnarray}
where the effective line tension $\lambda_{\rm eff}$ is defined as, 
\begin{eqnarray}
\lambda_{\rm eff}(x)=\lambda_0\,\left(1-\frac{\rm T}{\rm T_m}\right)
+\left(\frac{3Q{\rm k_BT}}{2n_0\,l}\right)\,x^{1/2}\:.
\label{leff}
\end{eqnarray}
As a result of the renormalization of Eq.(\ref{b1}) 
by the excluded area constraints due to other pores, the effective line 
tension becomes a function of the pore size. This is the origin of the 
packing stabilization mechanism that leads to sieve-like structure of membranes. 
Indeed, as already mentioned above, a single pore destabilizes the membrane 
at high temperature ${\rm T}\ge {\rm T_m}$ as the effective line tension 
becomes negative. In a multiple pore system, however, the membrane remains
stable even beyond $T_m$. For temperatures such that ${\rm
T_m}<{\rm T_c}\leq {\rm T}$, $G(x)$ in Eq.~(\ref{fener}) 
admits a minimum at $x=x_1$ such that
\begin{eqnarray}
\lambda_{\rm eff}(x_1)=0\Leftrightarrow\sqrt{x_1}&=&\left(\frac{Q_c}{Q}\right)
\left(\frac{\rm T_c}{\rm T}\right)
\left(\frac{\rm T-T_m}{\rm T_c-T_m}\right)\,,
\label{x1}
\end{eqnarray}
where $Q_c=Q({\rm T}={\rm T_c})$ and ${\rm T_c}$, the critical temperature
at which $x_1$ coincides with $x=1$, is given 
by $3Q_c=2Q_0(1-{\rm T_m}/{\rm T_c})$, i. e.,
\begin{eqnarray}\label{TC}
\lefteqn{
\frac{2Q_0(1-y)}{3}={\rm e}^{-y\beta_m F_0}} 
\\ \nonumber && \times
\int_{1}^{\infty}\!\!\!\!\! dx\,
\exp\left\{Q_0\left(1-y\right)\,\left[(x-1) \! - \!
\frac{2}{3}\,x^{3/2}\right]\right\}
\end{eqnarray}
with $Q_0=n_0\,\ln(z)$ and $y={\rm T_m}/{\rm T_c}\:$. Interestingly, 
Eq.(\ref{x1}) states that the sizes corresponding to a vanishing line tension 
is that of stable pores. 

\begin{figure}[tbp]
\includegraphics[width=0.35\textwidth]{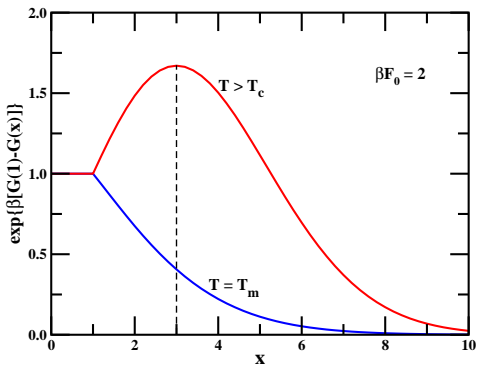} \caption{Reduced equilibrium
distribution, $P_{\rm eq}(x)/P_{\rm eq}(1)={\rm e}^{-\beta\left[
G(x)-G(1)\right]}$, of pore size $x$ for
$\beta F_0=2$, $Q_0=4$ and two temperature regimes. $G(x)$ is given 
in Eq.(\ref{fener}) with $Q=0.215$ for $T=T_m$ and $Q=8/(9\sqrt{3})=0.513$ 
for $T>T_c$ (i.e., for $T=3T_m/2$ and $T_c=1.082T_m$). The dashed vertical 
line indicates the maximum $x_1=3$ for $T>T_c$.} \label{fig2}
\end{figure}

\begin{figure}[tbp]
\includegraphics[width=0.35\textwidth]{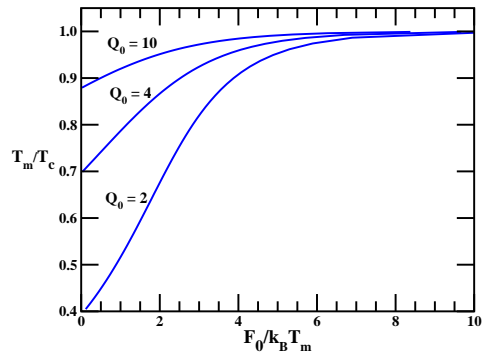} \caption{Ratio $T_m/T_c$ in
Eq.(\ref{TC}) as a function of the reduced energy of pore
formation.} \label{fig3}
\end{figure}

Two conclusions can be drawn from these results. First, the
excluded area constrains between pores can stabilize a membrane
even in parameter regions where the effective line tension of pores 
is negative. Second, the distribution of pore size in the
case of negative effective membrane line tension may have a
maximum at nonzero contour length as illustrated in Fig.~\ref{fig2}. 
Three regimes can be distinguished: (i) At low $T<T_m$, the line 
tension is positive, and the distribution $P_{\rm eq}(n)$ 
of pore size drops monotonically as a function of $n$. (ii) At
intermediate temperatures, $T_m < T < T_c$, the line tension is
negative, $P_{\rm eq}(n)$ still drops monotonically, but the pores 
are now stabilized due to the presence of the others. (iii)
At high $T>T_c$, a maximum emerges in $P_{\rm eq}(n)$, 
i. e., pores have a most probable size. Figure~\ref{fig2} shows the 
reduced distribution of pore size,
$P_{\rm eq}(x)/P_{\rm eq}(1)$, for two temperatures below and above
$T_c$, with a maximum at $x=3$ (or $n=3n_0$) for $T > T_c$. This 
regime sensitively depends on the bare pore free energy, $F_0$, 
required for creation of a minimal-sized pore. This is 
illustrated in Fig.~\ref{fig3} where ${\rm T_c}$ decreases towards 
${\rm T_m}$ (i.e., ${\rm T_m}/{\rm T_c}$ increases towards $1$) 
when increasing either $F_0$ or the effective smallest pore size $Q_0$.

From a practical point of view, the porous membrane state is 
interesting because it should have peculiar permeability characteristics.
Apart from the number and size of pores, another quantity
determining the permeability is the lifetime of pores, i. e.,
how long a pore stays open once created. We now discuss
briefly how the lifetime of pores is affected by the presence
of the other pores.

Six processes contribute to the dynamical evolution of the pores: 
Pore opening and closing, pore growth and shrinking, pore coalescence 
and splitting.  The pore opening and closing is mainly controlled 
by the potential barrier that must be overcome to create a pore: 
the amphiphiles must change their orientation, and the free energy 
in the intermediate state is different from the corresponding 
energy when a pore already exists. The characteristic time scale
of this process does not depend on the pore packing and shall not 
be considered here. Amphiphile rearrangements also take place 
when pores coalesce and split, hence these processes are 
rate-driven and the characteristic time scale depends on the
height of a potential barrier. For simplicity, we shall assume
that this height is very large, i. e., we shall neglect coalescence
and splitting events. Hence we are left with the characteristic
time $\tau$ for the growing and shrinking of a pore. Our main interest 
in this quantity is to allow comparison on the pore lifetime in 
two distinct situations: a single pore in a membrane versus a 
pore in a gas phase ensemble of pores in a membrane.

\begin{figure}[tbp]
\includegraphics[width=0.35\textwidth]{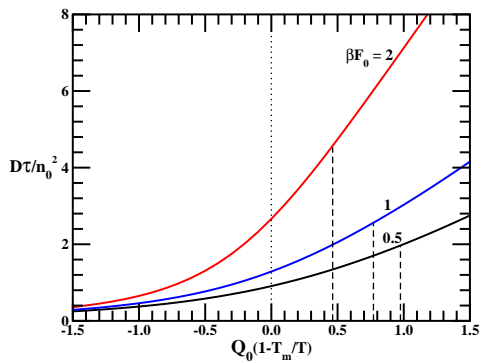} 
\caption{Reduced pore shrinking
time in Eq.(\ref{life}) as a function of the reduced temperature
$Q_0 (1-T_m/T)$ for different $\beta F_0$. The dotted vertical
line indicates $T=T_m$ above which the line tension becomes negative. 
Dashed vertical lines indicate the crossover points $T=T_c$ [i.e., 
$Q_0 (1-T_m/T_c)=0.975,\:0.771,\:0.463$ for $\beta F_0=0.5,\:1,\:2$, 
respectively] where $P_{\rm eq}(x)$ starts to exhibit a maximum like
in Fig.~\ref{fig3}.} \label{fig4}
\end{figure}

Neglecting hydrodynamic effects~\cite{fn_hydro}, the growth and 
shrinking dynamics of the pore can fairly be described by a diffusion 
process with a diffusion constant $D$ (assumed here independent of 
$x$) in the potential $G(x)$. Then, $\tau$ is given by the mean time 
that an already existing pore takes to first reach the minimum pore 
size $x=1$ (or, $n_0$). According to the first passage time theory, 
the lifetime $\tau$ can be estimated from the
relation~\cite{SSS80,Risken,BS97}
\begin{equation}\label{life}
\tau=\frac{n_0^2}{D}\,
\frac{\displaystyle{\int_{1}^{\infty}\!dx\,{\rm e}^{\beta G(x)}\,
\left[\int_{x}^{\infty}\!{\rm e}^{-\beta G(y)}\,dy\right]^2}}
{\displaystyle{\int_{1}^{\infty}\!{\rm e}^{-\beta G(x)}\,dx}}\:.
\end{equation}
In the case of a single pore in a membrane, Eq.(\ref{life}) reduces
to the pore lifetime $\tau_0$ as,
\begin{equation}\label{life1}
\tau_0=\frac{n_0^2}{D}\,\left[Q_0\left(\frac{\rm T_m}{\rm
T}-1\right)\right]^{-2}\:\:;\:\:{\rm T}<{\rm T_m}\:.
\end{equation}
This $\tau_0$ quadratically increases with temperature and diverges 
with the membrane disintegration as ${\rm T}$ approaches 
${\rm T_m}$. As displayed in Fig.~\ref{fig4}, in the case of an 
ensemble of pores, the pore lifetime $\tau$ begins to increase 
considerably at $T_m$ but without diverging and, thus with no 
membrane disintegration. At ${\rm T}$ below ${\rm T_c}$, a pore of 
any size will ultimately shrink to disappearance on average because 
of the drift towards $x=1$. In the stabilized regime above $T_c$, the 
pore size diffuses towards the minimum $x_1>1$ of $G(x)$, and $x=1$ 
can be reached {\em via} an escape process over the energy barrier, 
$\Delta=G(1)-G(x_1)$, such that $\tau$ scales as $\tau \propto \exp(\beta \:\Delta)$. 
As a result, the pore remains open much longer in the packing stabilized 
regime without membrane disintegration.

To summarize, we have analyzed the statistics of multiple pore
formation in membranes without surface tension within a classical 
van-der Waals approach. We have shown that the exclude area
interactions between pores allows membrane stabilization against 
disintegration even when the effective line tension becomes negative 
at high temperatures. In a certain range of parameters, this leads 
to a nano-porous membrane state where the membrane has a sieve 
structure with long-lived holes of finite size. Owing to the diversity 
of biological systems (all the more artificial lipid bilayers) and 
wide range of accessible parameters, we expect such stable multipore
membranes exist at physiological conditions with pore sizes 
in the range of $(1 - 10)\:nm$. This invites speculations on 
possible applications of such structures, e.g., membranes with 
selective permeability for controlled drug delivery, 
or to promote biomolecule translocation.

Our result illustrates the rich diversity of membrane structures
that can form by chemically or physically tuning the line tension 
$\lambda$. As a crude  approximation, we have 
$\lambda\propto \gamma_0 h$, where $\gamma_0$ is the 
free energy per unit area between coexisting regions of hydrophobic
and hydrophilic molecules, and $h$ the membrane thickness. Such an 
estimation is valid only for the so-called hydrophobic pores,
while for hydrophilic pores, the scaling is $\lambda\propto \kappa/h$, 
where $\kappa \propto h^2$ is the bending modulus and $1/h$ the 
membrane pore curvature. As $\lambda\propto h$ in any case, one would 
expect sieve-like membrane structures to be more likely in thin membranes 
although other mechanisms do exist, for example, in mixed membranes 
or membranes with additives, that lead to reduction of the line tension.


\begin{thebibliography}{10}
\bibitem{fn_membrane}
We assume that the membrane is still stable or at least metastable. Obviously,
  an alternative, competing state is a fluid of membrane patches. We assume
  that this state is suppressed for some reason - either kinetically or
  thermodynamically. For example, it will be prevented at high lipid
  concentration, since the lipids can then be packed more efficiently in a
  membrane stack (smectic phase) than in a fluid of aligned bilayer patches
  (nematic phase).
\bibitem{fn_hydro}
The hydrodynamic time scale $\tau_h \propto \eta h^2/\lambda $ ($\eta $ is the
  effective viscosity, including the membrane and the surrounding liquid, $h$
  is the membrane thickness, and $\lambda $ the line tension), which roughly
  demarcates the crossover from the diffusion regime to the hydrodynamic
  regime, is of the order of $(1 - 10^{2})\, s$ and thus presumably much larger
  than the typical life time (msec) of unstable or metastable pores.
\end{thebibliography}
\end{document}